\documentclass[aps,twocolumn,floatfix,superscriptaddress]{revtex4-1}
\usepackage[british]{babel}
\usepackage[utf8]{inputenc}
\usepackage{lmodern}
\usepackage[T1]{fontenc}
\usepackage{hyperref}

\usepackage{mathtools} 
\usepackage{amsfonts} 

\usepackage{graphicx}
\usepackage{dcolumn}
\usepackage{bm}
\usepackage{subcaption}
\usepackage{booktabs} 
\usepackage{tabularx} 
\usepackage{multirow} 
\usepackage[table]{xcolor} 

\DeclareMathOperator{\dw}{dw}

\begin{document}

\title{Numerical study of the roughness of domain walls in the 2-dimensional random bond Ising model using a weighed-loop algorithm}

\author{Rick Keesman}\affiliation{Department of Information and Computing Sciences,
	Utrecht University, P.O. Box 80089, 3508 TB Utrecht, The Netherlands}\affiliation{\textup{Corresponding author (\texttt{r.keesman@uu.nl})}}
\author{P. J. Overbeeke}\affiliation{Mathematical Institute, Utrecht University,
P.O. Box 80010, 3508 TA Utrecht, The Netherlands}

\date{\today}

\begin{abstract}
We introduce a weighed-loop algorithm that is applicable to any weighed graph network. It is designed to prefer a route of energetically unfavourable bonds in the lattice that can then be flipped without changing the structure inside and outside the enclosed loop. Due to this property there are effectively no energy barriers thus making this algorithm very suitable for finding low energy states in very rough energy landscapes. We apply this algorithm to the random bond Ising model with domain walls and show that the weighed-loop algorithm can outperform Niedermayer's algorithm for low enough temperatures and high enough disorder. We consolidate the high-temperature behaviour of the roughness of a domain-wall with a low-temperature expansion presented in this paper and show agreement with results from our simulations. The effects of temperature, disorder, and system size on the roughness of domain-walls is also investigated.
\end{abstract}

\maketitle

\section{Introduction}\label{sec:rbim1}
The behaviour of field and current-driven magnetic domain walls is of great interest due to the promising applications that domain walls have in magnetic memory devices and logic devices~\cite{PHT_08,AXF_05}. When the magnetic material contains impurities, that can either be created by accident during the manufacturing process or injected intentionally, the motion of the domain walls at low enough driving currents is dominated by the movement of segments of the wall between so-called pinning sites~\cite{MCA_07}. A critical field or current separates the creep regime as described above and the flow regime in which the driving force is large enough for linear behaviour~\cite{LFC_98}. Interestingly, this behaviour is found in a plethora of physical systems like propagating crack fronts~\cite{BBL_93} and moving vortices in superconductors\cite{BFG_94}.

In this paper we choose to study the square-lattice Ising model where disorder is introduced in the form of random bonds between nearest neighbouring sites. Although the two-dimensional Ising model provides a highly idealized playground it does contain most of the characteristics of thin films of magnetic material with perpendicular anisotropy (PMA). In both cases the individual spins are either (almost completely) up or down resulting in a very sharp domain wall such that the Ising model relates closely to thin-film PMA's.

The dynamics in the creep regime and static properties of domain walls are related through the roughness (or wandering) exponent. In case of domain walls in the 2D random-bond Ising model (RBIM) this exponent is $\zeta_{\text{RB}}=2/3$~\cite{HH_85,BFG_94} and is in agreement with experiments on PMA's~\cite{LFC_98,AWM_10,RMH_10,CRC_04}. In case of random-field disorder this exponent is $\zeta_{\text{RF}}=1$~\cite{Nat_97,DZZ_12} in contrast to $\zeta_{\text{T}}=1/2$ when thermal fluctuations dominate the behaviour. When both random-bond disorder and thermal fluctuations are considered a cross-over takes place at a typical length scale characterized by the Larkin length $L_c$~\cite{FGL_89} above which the random-bond disorder induced roughness dominates.

Here we set out to investigate the static properties of square-lattice RBIM domain walls by means of Monte Carlo simulations. Due to disorder the energy landscape becomes rough and in most cases results in highly degenerate ground states. Many conventional algorithms result in over-sampling parts of phase space due to the exponentially suppressed probability of leaving a certain meta-stable state. In the last few decades many new update algorithms have emerged~\cite{Nie_88,Lee_93,Hou_01,ZOK_15} as well as other simulation schemes~\cite{BBO_15,TM_16,SW_86} to overcome these obstacles. Although most loop algorithms follow local update schemes the big advantage is that they are not inhibited by energetic barriers between meta-stable states. We propose a new type of loop algorithm, a weighed-loop algorithm, where the loop is not constructed by a random walk but instead allows for different possible routes to have different weights assigned as to improve the acceptance probability of suggested updates.

The outline of this paper is as follows. First we discuss the model in more detail in section~\ref{sec:rbim2} and combine previously obtained theoretical results for the static properties of domain walls together with our low-disorder low-temperature expansion. In section~\ref{sec:rbim3} we present the weighed-loop algorithm, prove detailed balance and ergodicity, and discuss some other properties. We also compare the weighed-loop algorithm to Niedermayer's algorithm~\cite{Nie_88} and show that there exists a region in the parameter space consisting at low temperature and high disorder where the weighed-loop algorithm is decorrelates faster. In section~\ref{sec:rbim4} we present the results and the analysis of our simulations on static properties of the domain walls after which we conclude with a discussion and conclusion in section~\ref{sec:rbim5}.

\section{Model and Theory}\label{sec:rbim2}
In the standard Ising model each lattice point $i$ is occupied by a spin up $\sigma_i=+1$ or a spin down $\sigma_i=-1$ particle. The Hamiltonian is defined by
\begin{align}\label{eq:hamiltoniandef}
	\mathcal{H}(\sigma)=-\frac{1}{2}\sum_{i\in\Lambda} \sum_{j\in\mathcal{N}_i}J_{ij}\sigma_i\sigma_j \, ,
\end{align}
where $\Lambda$ is the entire lattice, $\mathcal{N}_i$ are the neighbouring spins of spin $i$ and $J_{ij}$ the bond strength between spins $i$ and $j$. Using a $L\times H$ size rectangular lattice and using boundary conditions with a periodic boundary along the $L$-direction and an anti-periodic boundary along the $H$-direction, the lattice is topological homeomorph to a Klein bottle. Therefore the lattice induces at least one domain wall. Throughout the paper we assume ferromagnetic bonds of varying strengths; $J_{ij}=J\pm J\Delta$ with $0\leq\Delta\leq1$. The disorder is local and uncorrelated between bonds such that the disorder averages are given by $\langle J_{i j} \rangle_{\text{dis}}=J$ and $\langle J_{i j} J_{i' j'} \rangle_{\text{dis}}=J^2+J^2\Delta^2 \delta_{i i'} \delta_{j j'}$.

The horizontal displacement of a domain wall at height $h$ for a realized configuration is given by the function $\dw(h)=\sum_{l=1}^{L}\sigma_{l h}$, where $0\leq h<H$ and $(l,h)$ is the Cartesian coordinate for the position of spin $\sigma_{l h}$. Below the critical temperature the probability that the wall is as wide as it is high is negligible. To ensure that the horizontal displacement is calculated correctly the anti-periodic boundary can be shifted such that it does not cross the domain wall. This elementary definition allows for a quick computation of the position of domain walls at the cost of a loss of accuracy due to isolated pockets of reversed spins. By definition overhangs are also not taken into account although these are ignored in derivations for roughness exponents as well. Both isolated pockets of spins and overhangs are energetically expensive excitations and so disappear for low temperatures and disorder. 

\begin{figure}[h]\centering
 	\includegraphics[width=0.75\columnwidth]{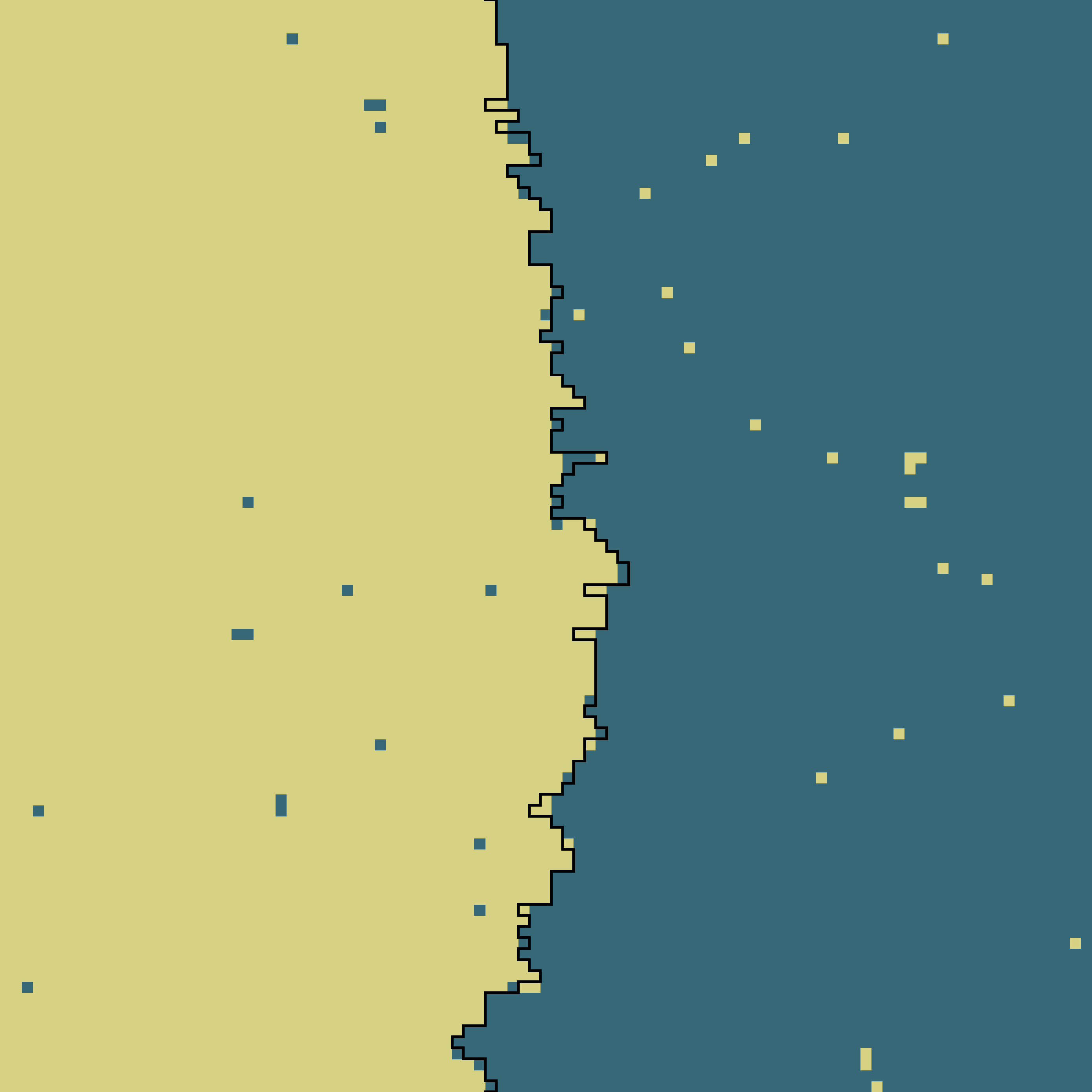}
	\caption{A typical spin configuration of a system of size $100\times100$ at $\beta J =0.7$ in the absence of disorder is shown. The system has a region of spins up (blue) and a region of spins down (yellow) separated by a domain wall. The horizontal displacement of the domain wall, as defined by $\dw(h)=\sum_{l=1}^{L}\sigma_{l h}$, is shown as a solid black line.}
	\label{fig:typicalWall}
\end{figure}

The system is in thermal contact with a bath at temperature $T$. We define the inverse temperature $\beta:=1/(k_B T)$ with $k_B$ the Boltzmann constant. We start each simulation with exactly one domain wall. Since extra pairs of domain walls cost macroscopic amounts of energy we are practically ensured to always have exactly one domain wall below the critical temperature $T_c$. An example of a typical spin configuration and the resulting horizontal displacement of the domain wall for a system of size $100\times100$ at $\beta J = 0.7$ in the absence of disorder is shown in Fig.~\ref{fig:typicalWall}.

To study the static properties of domain walls in more detail we consider the discrete Fourier transform of the horizontal displacement
\begin{align}\label{eq:fourierdef}
\mathcal{F}(p)&=\frac{1}{H}\sum_{h=0}^{H-1}\exp{\left(\frac{2\pi i p h}{H}\right)}\dw(h)\,,
\end{align}
and the structure factor $A(p,H,\beta J,\Delta)=\left \langle |\mathcal{F}(p)|^2 \right \rangle$ for modes $0<p<H-1$ as a thermal and disorder average over all possible spin configurations. We will show that without disorder $A(p,H,\beta J,\Delta=0)=A(H,\beta J)\csc^2{\left(\pi p/H\right)}$ where $A(H,\beta J)$ is a scaling factor. For a system without disorder at high temperatures the scaling factor $A_{\text{high}}(H,\beta J)$ is given by
\begin{align}\label{eq:ahigh}
A_{\text{high}}(L,\beta J)\sim \sinh^{-1}{\left(2\beta J + \ln{\tanh{\beta J}}\right)}\,,
\end{align}
which follows from capillary-wave theory~\cite{FFW_82} and results for the surface tension of the 2D Ising model~\cite{Ons_44}.

The next step is to make a low-temperature expansion for large domain walls in a system without disorder which will yield a low-temperature scaling factor $A_{\text{low}}(H,\beta J)$. We then combine this with the high-temperature scaling factor $A_{\text{high}}(H,\beta J)$ into $A(H,\beta J)$ that works for all temperatures.

\emph{Low-temperature expansion.} Consider the ground state of a system without disorder. The domain wall is then given by a straight wall without any perturbations which do not contribute to the structure factor. Let $A_d(p,H,\beta J)$ be the average value of $A(p,H,\beta J)$ of all systems with exactly $d$ defects. A defect is defined as a single horizontal displacement over a certain length $b$. Without affecting the non-zero modes we can assume that all defects occur at only one side of the domain wall without a loss of generality. Similarly, since we are eventually interested in the absolute norm of the Fourier components we assume that defects start at position $h=0$. For a domain wall with a single defect the horizontal displacement function is given by $\dw(h)=\theta(h-b)$ with $\theta(x)$ the step function. Using Eq.~\eqref{eq:fourierdef} this results in
\begin{align}\label{eq:onedef1}
	|\mathcal{F}(p)|^2&=\frac{1}{H^2}\csc^2{\left(\frac{\pi p}{H}\right)}\sin^2{\left(\frac{\pi b p}{H}\right)}\,.
\end{align}
Averaging over all non-trivial values of $b$ and taking the large $H$ limit while $0<p<(L-1)/2$ yields
\begin{align}\label{eq:a1}
A_1(p,H,\beta J)=\frac{1}{2H^2}\csc^2{\left(\frac{\pi p}{H}\right)}
\end{align}
At low temperatures and for large systems we can safely assume that the defects are non-interacting such that $A_d(p,H,\beta J)=d A_1(p,H,\beta J)$. We can then sum over all possible number of defects to get the correct expression for the structure factor
\begin{align}\label{eq:ap}
A(p,H,\beta J)=\sum_{d=0}^\infty A_d(p,H,\beta J)P(d, H, \beta J)\,,
\end{align}
where $P(d,H,\beta J)$ is the probability of finding $d$ defects.

Each defect increases the energy of the system by $4J$ and so the probability of finding $d$ defects can be rewritten as  $P(d,H,\beta J)=Z^{-1}(H,\beta J)g(d, H)\exp\left(-4d \beta J\right)$ with $Z$ the normalizing constant (or partition function) and $g(d,H)$ the number of possible ways $d$ defects can occur in a system of height $H$. We approximate $g(d,H)$ by assuming defects are still uncorrelated and so in the large $H$ limit $g(d,H)\approx \binom{H}{d}^2\approx\left(H^d/d!\right)^2$.
Note that this is an overestimation as destructive interference between defects is not taken into account. The probability of finding $d$ defects is then given by
\begin{align}\label{eq:prob}
P(d,H,\beta J)= I_0^{-1}\left( 2 H e^{-2 \beta J} \right)\left(\frac{H^d}{d!}\right)^2 e^{-4d \beta J}\, ,
\end{align}
where $I_n(x)$ denotes the Bessel function of the first kind. Using Eqs.~\eqref{eq:a1}-\eqref{eq:prob} this results in low temperature scaling factor
\begin{align}\label{eq:alow}
A_{\text{low}}(H,\beta J)=\frac{e^{-2\beta J}}{2H}\frac{I_1(2H e^{-2\beta J})}{I_0(2H e^{-2\beta J})}\,.
\end{align}

\emph{Scaling factor.}  To provide a single expression for the scaling factor we now naively combine the results at high temperatures~\cite{FFW_82} with the low-temperature expansion Eq.~\eqref{eq:alow} such that the scaling factor $A(H,\beta J)$ behaves similarly to the approximations in their respective temperature regimes. First let $a(H,\beta J)$ be the unknown pre-factor of $A_{\text{high}}(H,\beta J)$ in Eq.~\eqref{eq:ahigh}. For low temperatures $\beta_c \ll \beta $ we have $A_{\text{high}}(H,\beta J)\approx 2a(H,\beta J)e^{-2\beta J}$. Using the approximation $I_n(z)\approx\left(z/2\right)^n/\Gamma(n+1)$ for small $z$ results in the approximation $A_{\text{low}}(H,\beta J)\approx e^{-4\beta J}/2$. Similarly just below the critical point $0<\beta J-\beta_c J\ll 1$ we have $A_{\text{high}}(H,\beta J)\approx a(H,\beta J)/4 \left(\beta J-\beta_c J\right)^{-1}$. For large $z$ we have $I_n(z)\approx e^z /\sqrt{2\pi z}$ such that $A_{\text{low}}(H,\beta J)\approx e^{-2\beta J}/(2H)$. When $a(H,\beta J)=1/(4H)$ the low-temperature expansion of $A_{\text{high}}$ and high-temperature expansion of $A_{\text{low}}$ are connected. An elegant solution that fits both low- and high-temperature expansions of the scaling factor is then given by
\begin{align}\label{eq:a}
A(H,\beta J)=A_{\text{high}}(H,\beta J)\frac{I_1\left(4H^2 A_{\text{high}}(H,\beta J)\right)}{I_0\left(4H^2 A_{\text{high}}(H,\beta J)\right)}\,,
\end{align}
with 
\begin{align}\label{eq:ahigh2}
A_{\text{high}}(H,\beta J)=\frac{1}{4H} \sinh^{-1}{\left(2\beta J + \ln{\tanh{\beta J}}\right)}\,.
\end{align}

\section{Weighed-Loop Algorithm}\label{sec:rbim3}
\begin{figure}[!htb]
	\centering
	\includegraphics[width=0.9\columnwidth]{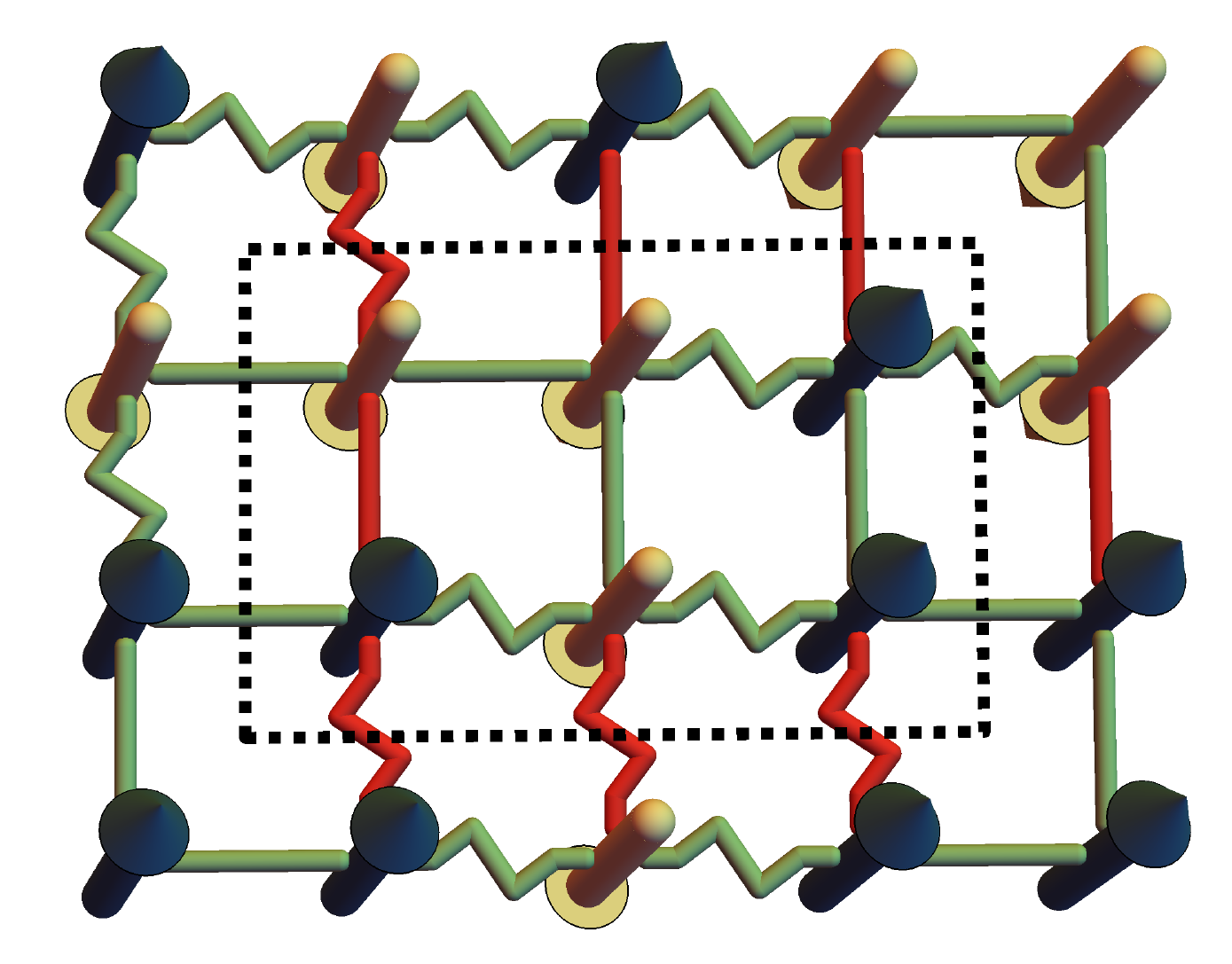}
	\caption{Part of a typical RBIM spin configuration with spins pointing up (blue) or down (yellow). The lines between neighbouring spins are ferromagnetic (straight) or anti-ferromagnetic (wiggled). Bonds that are satisfied, that is to say they contribute negatively to the Hamiltonian, are coloured green whereas unsatisfied bonds are coloured red. Flipping any individual spin increases the energy but flipping the cluster of spins inside the black dashed line (loop) yields a decrease in energy. Although in our model anti-ferromagnetic bonds are strictly not allowed ($\Delta \leq 1$) the argument still holds for the combination of strong and weak ferromagnetic bonds.}
	\label{fig:loop}
\end{figure}

The Ising model has two-fold degenerate ground states that become highly degenerate when disorder is introduced in large enough systems. The energy landscape becomes rough and at low temperatures many algorithms get stuck in metastable states surrounded by energy barriers that become exponentially higher as temperature is lowered. Many single-spin-flip and cluster-growing algorithms suffer from this since they are dependent on all of the bonds around and inside the cluster. Although loop algorithms are local update algorithms the main advantage is that they only depend on the bonds that the loop follows and not on the bonds inside the surrounded cluster. A part of a configuration is shown in Fig.~\ref{fig:loop} that illustrates this point. For this example let the bonds be ferromagnetic (straight bond) or antiferromagnetic (wiggled bond). Each spin has three or more energetically favourable bonds (green) and so flipping any individual spin always increases energy. The dashed line shows a cluster of spins that, if flipped all together, will lower the energy. The main idea behind the weighed-loop algorithm is that bonds are iteratively added to loop to form a closed chain of bonds with a bias toward bonds that are energetically favourable to swap. Here we will explain the algorithm in more detail, prove detailed balance and ergodicity, and compare the autocorrelation times for domain-walls with those from Niedermayer's algorithm.

\emph{Details algorithm.} Let $G$ be the graph given by the lattice $\Lambda$ such that the spins are the faces of $G$. The algorithm first selects a starting node of $G$ at random. Given that the algorithm cannot go backwards each step after the first the algorithm has three possible edges of $G$ to add to the current chain. Each edge corresponds to a bond $J_{ij}$ and the algorithm adds the bond with a relative weight $W(J_{ij})=\exp(-\beta J_{ij}\sigma_i\sigma_j)$. Although any weight can be assigned to any of the bonds we choose these particular values as the relative probability between two bonds that can be followed by the weighed-loop algorithm equal to the relative local Boltzmann weights of the bonds. Available bonds with equal energy have equal probability of being chosen and at lower temperatures bonds that will lower the energy more than other bonds have a higher probability. The algorithm continues adding edges of $G$ until a closed cycle in $G$ is constructed. The spins enclosed by the closed cycle define the cluster to flip. Since the lattice is homeomorph to a Klein bottle it is possible for the closed cycle to enclose the entire lattice. If such a loop is constructed, an additional loop is constructed. If both loops enclose the entire lattice, the spins enclosed by both loops define the cluster to flip. The algorithm will almost always construct a closed loop with an additional tail. Since the weights remain unchanged in the tail after the update it is used as part of the new loop.

Given that flipping the cluster of spins defined by the loop $\lambda$ transitions the system from $\mu$ to $\nu$, we define the acceptance probability of $\lambda$ as
\begin{align}
	\label{eq:acceptance}
	A_{\mu\rightarrow\nu}(\lambda)=\min{\left(1,e^{-\beta\left(E_{\nu}-E_{\mu}\right)}\frac{g_\nu(\lambda)}{g_\mu(\lambda)}\right)}\,.
\end{align}
In the above equation $g_\mu(\lambda)$ is the probability of selecting $\lambda$ in $\mu$ which is defined by the product of the individual stochastic choices of the chaining of the bonds.

\emph{Properties.} To prove that with the acceptance probability~\eqref{eq:acceptance} the weighed-loop algorithm satisfies the detailed balance equation let $\Theta_{\mu\rightarrow\nu}$ be the set of all loops which transition $\mu$ to $\nu$ and let $\Theta_{\nu\rightarrow\mu}$ be the set of all loops which transition $\nu$ to $\mu$. The transitions probability are
\begin{align}
	\Pi(\mu\rightarrow\nu)=\sum_{\lambda\in\Theta_{\mu\rightarrow\nu}}g_\mu(\lambda)A_{\mu\rightarrow\nu}(\lambda)\,,\\
	\Pi(\nu\rightarrow\mu)=\sum_{\lambda\in\Theta_{\nu\rightarrow\mu}}g_\nu(\lambda)A_{\nu\rightarrow\mu}(\lambda)\,.
\end{align}
Each loop in $\Theta_{\mu\rightarrow\nu}$ is also a loop in $\Theta_{\nu\rightarrow\mu}$ hence summing over $\Theta_{\mu\rightarrow\nu}$ yields the same result as summing over $\Theta_{\nu\rightarrow\mu}$. Using Eq.~\eqref{eq:acceptance} and the equilibrium distribution given by the Boltzmann weights $\pi(\mu):=\exp(-\beta E_\mu)$ one can show that detailed balance $\pi(\mu)\Pi(\mu\rightarrow\nu)=\pi(\nu)\Pi(\nu\rightarrow\mu)$ holds.

To prove ergodicity of the weighed-loop algorithm it suffices to show it can perform any single spin flip at finite temperature. The starting node of $G$ is chosen at random from a flat distribution for all possibilities. The weight $W(J_{ij})$ for any edge is non-zero and so there is always a finite probability that a loop around a single spin is constructed. The resulting acceptance probability in Eq.~\eqref{eq:acceptance} is, by construction, non-zero as well and so there is a non-zero probability the proposed loop is accepted and the single spin inside it flipped. Thus the weighed-loop algorithm is ergodic.

Although we made the choice that the relative weights of possible edges only depends on the energy difference one would get after flipping the bonds many other extensions are possible. The weights can be any type of function and it might be interesting to see the effects of a simulations with relative or absolute directional preferences. By doubling the relative weight of the edge in the forwards direction for example, the area enveloped by suggested loops would on average be larger. 

\emph{Comparison with Niedermayer's algorithm.} 
We compare the weighed-loop algorithm to Niedermayer's algorithm~\cite{Nie_88}. Niedermayer's algorithm can be viewed as an extension of the Swendsen-Wang~\cite{SW_87} and Wolff~\cite{Wol_89} algorithms that can be applied to glassy spin systems. Niedermayer's algorithms is almost the same as Wolff's algorithm except that any spin can be added to the cluster with varying probability. For Niedermayer's algorithm, with $E_{ij}:=\sigma_i\sigma_J J_{ij}$, the acceptance probability is given by
\begin{align}
P_{\text{acc}}=\begin{cases} 1-e^{-\beta (E_{ij}-E_0)}, & \mbox{if } E_{ij}\leq E_0 \\ 0, & \mbox{otherwise}\,, \end{cases}
\end{align}
with $E_0$ a free to choose parameter. If $E_0$ is larger than all possible $E_{ij}$ the acceptance probability for flipping the cluster is always $1$. Unfortunately, there is not much known about the effect of the $E_0$ on the correlation times~\cite{NB_99}. We have set $E_0=J+\Delta J$ so that the probability for flipping a selected cluster is always equal to unity.

\begin{figure}
	\includegraphics[width=0.95\linewidth]{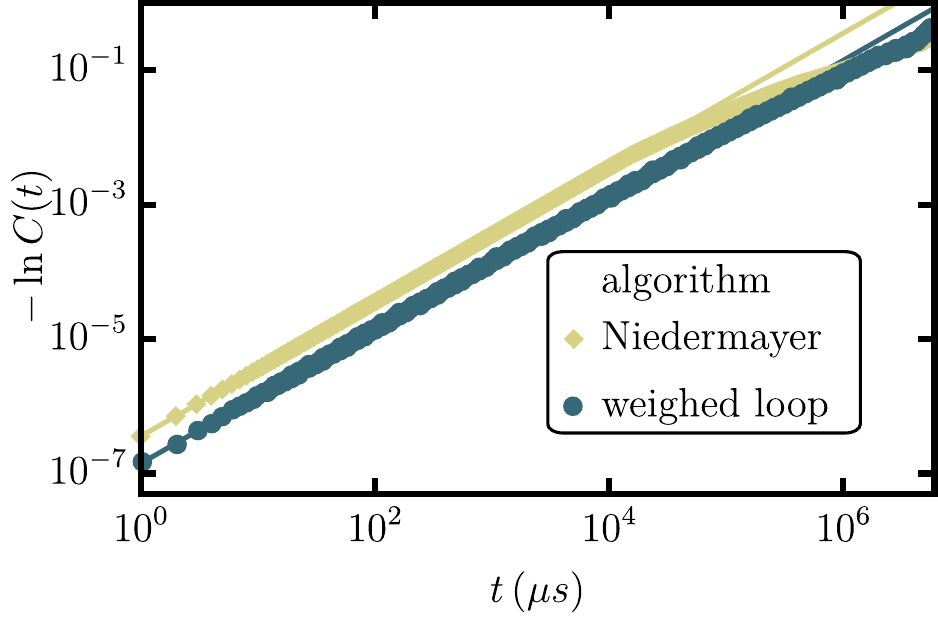}
	\caption{The negative logarithm of the autocorrelation of the first mode of the fourier coefficients $C(t)$ for both Niedermayer's algorithm (yellow diamonds) and the weighed-loop algorithm (blue disks) is shown for parameters $H=128$, $\beta=1$ and $\Delta=0.5$ as a function of real computation time $t$ in microseconds. The decorrelation happens exponentially at first after which a slowing down occurs. Best unweighed exponential fits up to $t=10^3$ are shown as solid lines in the corresponding colours.}
	\label{fig:compareExample}
\end{figure}

For both Niedermayer's algorithm and the weighed-loop algorithm we ran multiple simulations, each exactly $10$ minutes on the same system, on the same $24$ different disorder configurations for varying $H$, $\beta$, and $\Delta$. We measured the autocorrelation time of the squared amplitude of the first mode of the domain wall  $C(t):=\left\langle \left(|\mathcal{F}(1,t)|^2 - \left\langle|\mathcal{F}(1)|^2\right\rangle\right)\left(|\mathcal{F}(1,0)|^2 - \left\langle|\mathcal{F}(1)|^2\right\rangle\right)\right\rangle$ for both algorithms. Here $\mathcal{F}(1,t)$ is the first Fourier mode, the mode with the largest correlation time, at time $t$ and brackets denote the disorder and thermal average. For small $t$ we can assume exponential decay $C(t)\sim\exp(-a t)$ with equal pre factors for both algorithms. Figure~\ref{fig:compareExample} shows typical autocorrelation times for both algorithms evaluated on the same disorder configuration. We measure $a$ for both algorithms for varying parameters $H$, $\beta$, and $\Delta$ where larger values of $a>0$ lead to faster decorrelation. The results are listed in Table~\ref{tab:compareNieder}. We see a clear trend that the weighed-loop algorithm has faster decreasing correlations in the parameter regime where conventional algorithms tend to get stuck (low temperatures and high disorder). Additionally, on larger time scales as can be seen in Fig.~\ref{fig:compareExample}, we observe that the weighed-loop algorithm also decorrelates exponentially on longer time scales. This might be an indication that the weighed-loop algorithm tends to hop more frequently between far-removed local minima.

\begin{table}

	\begin{tabularx}{0.9\linewidth}{*3X} \toprule
		$\Delta$ & Niedermayer ($a$) & weighed-loop ($a$) \\	
		\bottomrule\hline
				 $H=32$, $\beta=1$ & & \\
		\hline
				$0.1$ & $\cellcolor{lightgray}6.5(2)\times10^{-4}$ & $1.35(4)\times10^{-5}$  \\
				$0.3$ & $\cellcolor{lightgray}1.8(6)\times10^{-5}$ & $1.42(8)\times10^{-5}$ \\
				$0.5$ & $\cellcolor{lightgray}2.1(3)\times10^{-5}$ & $1.3(2)\times10^{-5}$ \\
		\bottomrule
		\hline
				 $H=64$, $\beta=1$ & & \\
		\hline
				$0.1$ & $\cellcolor{lightgray}9.4(9)\times10^{-7}$ & $8.7(6)\times10^{-7}$  \\
				$0.3$ & $\cellcolor{lightgray}1.4(1)\times10^{-6}$ & $8.9(8)\times10^{-7}$ \\
				$0.5$ & $1.0(1)\times10^{-6}$ & $\cellcolor{lightgray}1.2(2)\times10^{-6}$ \\
		\bottomrule
		\hline
				 $H=128$, $\beta=1$ & & \\
		\hline
				$0.1$ & $\cellcolor{lightgray}2.0(2)\times10^{-7}$ & $4.5(4)\times10^{-8}$  \\
				$0.3$ & $\cellcolor{lightgray}5.3(6)\times10^{-7}$ & $8.1(7)\times10^{-8}$ \\
				$0.5$ & $\cellcolor{lightgray}6.4(5)\times10^{-7}$ & $1.1(2)\times10^{-7}$ \\
		\bottomrule
		\hline
				 $H=32$, $\beta=2$ & & \\
		\hline
				$0.1$ & $\cellcolor{lightgray}7.9(4)\times10^{-5}$ & $9.4(6)\times10^{-6}$  \\
				$0.3$ & $1.1(2)\times10^{-6}$ & $\cellcolor{lightgray}1.1(1)\times10^{-5}$ \\
				$0.5$ & $0.5(1)\times10^{-6}$ & $\cellcolor{lightgray}1.3(3)\times10^{-5}$ \\
		\bottomrule
		\hline
				 $H=64$, $\beta=2$ & & \\
		\hline
				$0.1$ & $2.6(3)\times10^{-7}$ & $\cellcolor{lightgray}8.4(9)\times10^{-7}$  \\
				$0.3$ & $4.7(5)\times10^{-7}$ & $\cellcolor{lightgray}1.2(2)\times10^{-6}$ \\
				$0.5$ & $1.9(2)\times10^{-7}$ & $\cellcolor{lightgray}1.8(5)\times10^{-6}$ \\
		\bottomrule
		\hline
				 $H=128$, $\beta=2$ & & \\
		\hline
				$0.1$ & $\cellcolor{lightgray}0.4(2)\times10^{-6}$ & $6.1(8)\times10^{-8}$  \\
				$0.3$ & $\cellcolor{lightgray}1.9(3)\times10^{-7}$ & $1.3(2)\times10^{-7}$ \\
				$0.5$ & $8.7(9)\times10^{-8}$ & $\cellcolor{lightgray}2.5(3)\times10^{-7}$ \\
		\bottomrule
		\hline
	\end{tabularx}
	\caption{For both the Niedermayer as well as the weighed-loop algorithm $a$ has been measured for a variety of parameter values. At each parameter point the greatest value, and thus the algorithm with faster decorrelation, has a light grey background to visualize the trend that the weighed-loop algorithm starts to outperform Niedermayer's algorithm for low enough temperatures and high enough disorder. The best values for $a$ come from averages of the best fits to the first $120$ data points obtained from simulations similar as shown in Fig.~\ref{fig:compareExample}.}
	\label{tab:compareNieder}
\end{table}

\section{Results}\label{sec:rbim4}
In this section we present several results obtained using the weighed-loop algorithm. First we compare simulations in the absence of disorder with the theoretical result Eqs.~(\ref{eq:a},\ref{eq:ahigh2}) obtained from our low-temperature expansion in combination with the high-temperature expression resulting from the surface tension of the 2D Ising model~\cite{Ons_44} and capillary-wave theory~\cite{FFW_82}. Next we induce disorder in the system and measure the typical length $L_c$ at which the system switches from a roughness dominated by thermal fluctuations to higher length scales at which the roughness is determined by disorder. 

\begin{figure} \centering
	\includegraphics[width=0.95\linewidth]{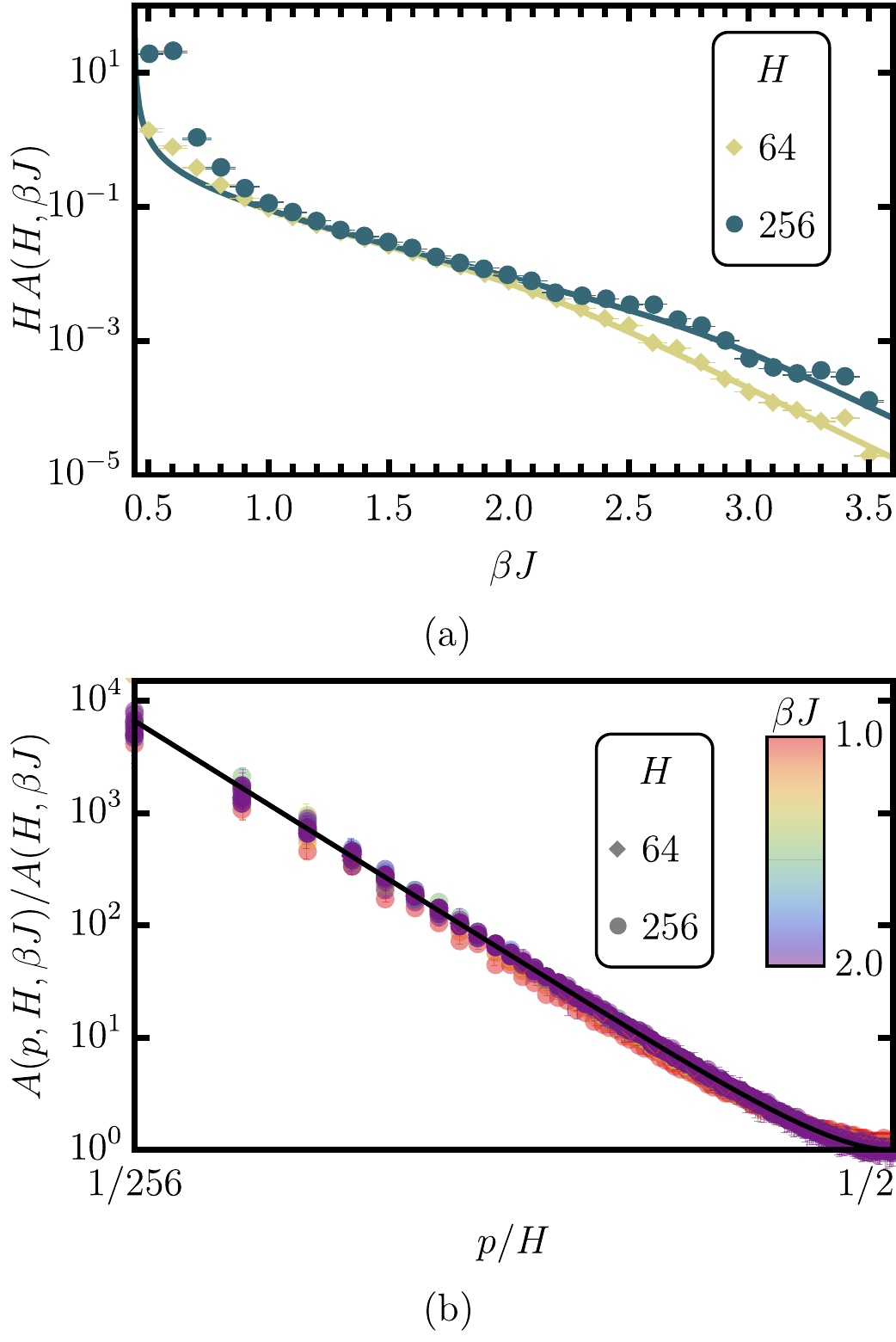}
	\caption{In (a) best values for $H A(H,\beta J)$ are shown as functions of $\beta J$ for systems of height $H=64$ (yellow) and $H=256$ (blue) together with Eqs.~(\ref{eq:a},\ref{eq:ahigh2}) shown as solid lines. These best values for $A(H,\beta J)$ come from best fits of $A(p,H,\beta J)$ which are shown as functions of $p/H$ in (b) for varying heights and moderate values for the temperature. The pre factor is scaled out to collapse the data onto $\csc^2(\pi p/H)$ shown as the solid black line. The error bars denote the random error margins which, at very high or very low temperatures, are negligible compared to the systematic errors as discussed in the text.}
	\label{fig:theoreticalWall}
\end{figure}

\emph{In the absence of disorder.} First we look at the simplest case $\Delta=0$ for which only thermal averaging is needed. The scaling factor $A(H,\beta J)$ for both $H=64$ and $H=256$ is shown as a function of temperature in Fig.~\ref{fig:theoreticalWall}(a). The data is in agreement with theory over the complete temperature-range that was used in the simulations. Error margins close to the critical temperature, at low values for $\beta J$, are much higher due to larger thermal fluctuations. This is also the region in which overhangs and local spin pockets start to occur. All these phenomena result in a less accurate approximation of structure of the domain wall. For extremely low temperatures the domain walls are mostly perfectly straight and so the error margins are increased due to the low probability for the wall to be in an excited state. Figure~\ref{fig:theoreticalWall}(b) shows a data collapse of the scaled mode-dependent factor $A(p,H,\beta J)/A(H,\beta J)$ as a function of $p/H$ to show the universal behaviour in the absence of disorder. The collapse indeed works best for moderate temperatures. 

\emph{Introducing disorder.} The addition of randomness to the system brings with a couple of complications. Besides the usual thermal averaging one also needs to perform disorder averaging resulting in much longer simulation runs than is the case without disorder. Additionally, the energy landscape becomes rough and so many more pockets of local energetically favourable configurations far away removed in phase space need to be explored to measure averages accurately. A trade-off is to be made between simulating large systems with faster dynamics and easier averaging at higher temperatures and less disorder or simulating smaller system in the more problematic regime of the $\Delta$-$\beta$ parameter space.

When disorder is introduced in an infinite system there is a typical length scale, known as the Larkin length $L_c$~\cite{FGL_89}, above which the roughness of the wall is determined by the disorder instead of thermal fluctuations. The roughness exponent from thermal fluctuations $\zeta_{\text{T}}=1/2$ differs from the exponent for random-bond disorder, for one-dimensional walls, $\zeta_{\text{RB}}=2/3$~\cite{HH_85,BFG_94}. The roughness exponent also appears in the average of the fourier modes. One finds $A(p,H,\beta J,\Delta) \sim p^{-(1+2\zeta)}$ in the low-mode limit, which is in agreement with $A(p,H,\beta J,\Delta) \sim \csc^2(\pi p / H)$ and $\zeta_{\text{T}}=1/2$ in the absence of disorder. The Larkin length can be translated into a typical mode $p_c/H:=1/(2 L_c)$ at which $\zeta$ changes from thermal to disorder-induced. Since the transition is relatively sharp we can make a fit for all $p$ of the form 
\begin{align}\label{eq:ap}
&A(p,H,\beta J,\Delta)=A(H,\beta J,\Delta) \times \nonumber\\
&			 			\left[ \{c(H,\beta J,\Delta) x^{-7/3} \}^b + \csc^{2b}(\pi p/H) \right ]^{1/b}\,. 
\end{align}
Here $A(H,\beta J,\Delta)$, $b$, and $c(H,\beta J,\Delta)$ are free fit-parameters from which $p_c$ can be extracted. For large systems and high enough temperatures we can scale out the height dependence $A(H,\beta J,\Delta)\sim1/H$ as can be seen from Fig.~\ref{fig:theoreticalWall}(a). 

\begin{figure}\centering
	\includegraphics[width=0.95\linewidth]{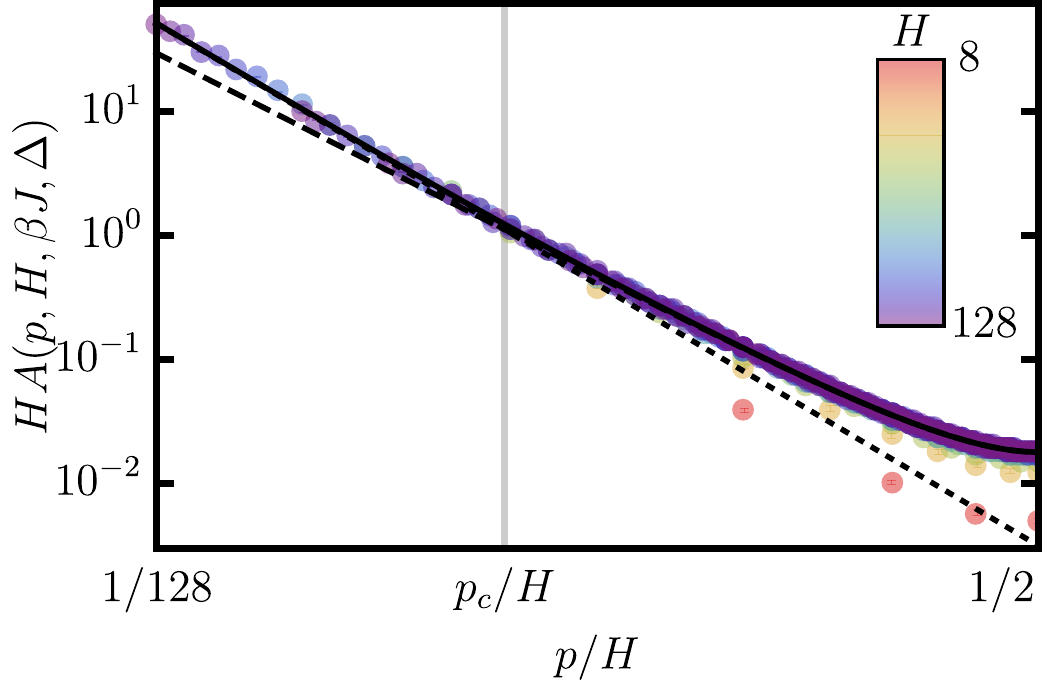}
	\caption{The average of the norm squared of the fourier modes $A(p,H,\beta J,\Delta)$ is plotted as a function of $p/H$ at fixed $\beta=2$ and $\Delta=0.15$ for different sizes of $8\leq H\leq128$. The best unweighed fit following Eq.~\ref{eq:ap} is shown as a solid black line and the power-law behaviour in the low-mode and high-mode regime are shown as dashed and dotted black lines respectively. From this best fit we find that the transition between these regimes happens at roughly $p_c\approx0.04H$, denoted by the vertical gray line, or equivalently $L_c\approx12.5$ in units of lattice spacing. For this fit only systems of size $H\geq48$ were taken into account as for smaller system sizes finite-size effects were too prominent.}\label{fig:varyL}
\end{figure}

In Fig.~\ref{fig:varyL} we look at $A(p,H,\beta J,\Delta)$ as a function of $p$ while varying the height $H$ of the system and keeping $\beta J=2.0$ and $\Delta=0.15$ fixed. Simulations were performed for $H=8,16,24,\dots,128$. For very small systems  $A(H,\beta J,\Delta)\sim1/H$ does no longer hold and additionally simulations of such systems does not yield data above the Larkin length $L_c$. Therefore, the unweighed fit of the form in Eq.~\eqref{eq:ap} is performed on data from systems with $H\geq48$ for which the higher-mode portion already sufficiently collapses. The figure also shows the best unweighed fit as a solid black line as well as a dashed and dotted line to visualize the power-law behaviour in the different regimes. We find a transition at $p_c\approx0.04H$ or equivalently $L_c\approx12.5$ in units of lattice spacing. Clearly one can extract the transition from a series of simulations of systems of slightly varying heights larger than the Larkin length rather well even for systems with disorder. We also find that $A(H,\beta J,\Delta)\sim1/H$ still holds, in the large-system limit at low enough temperatures, even when disorder is introduced to the system.

\begin{figure}\centering
	\includegraphics[width=0.95\linewidth]{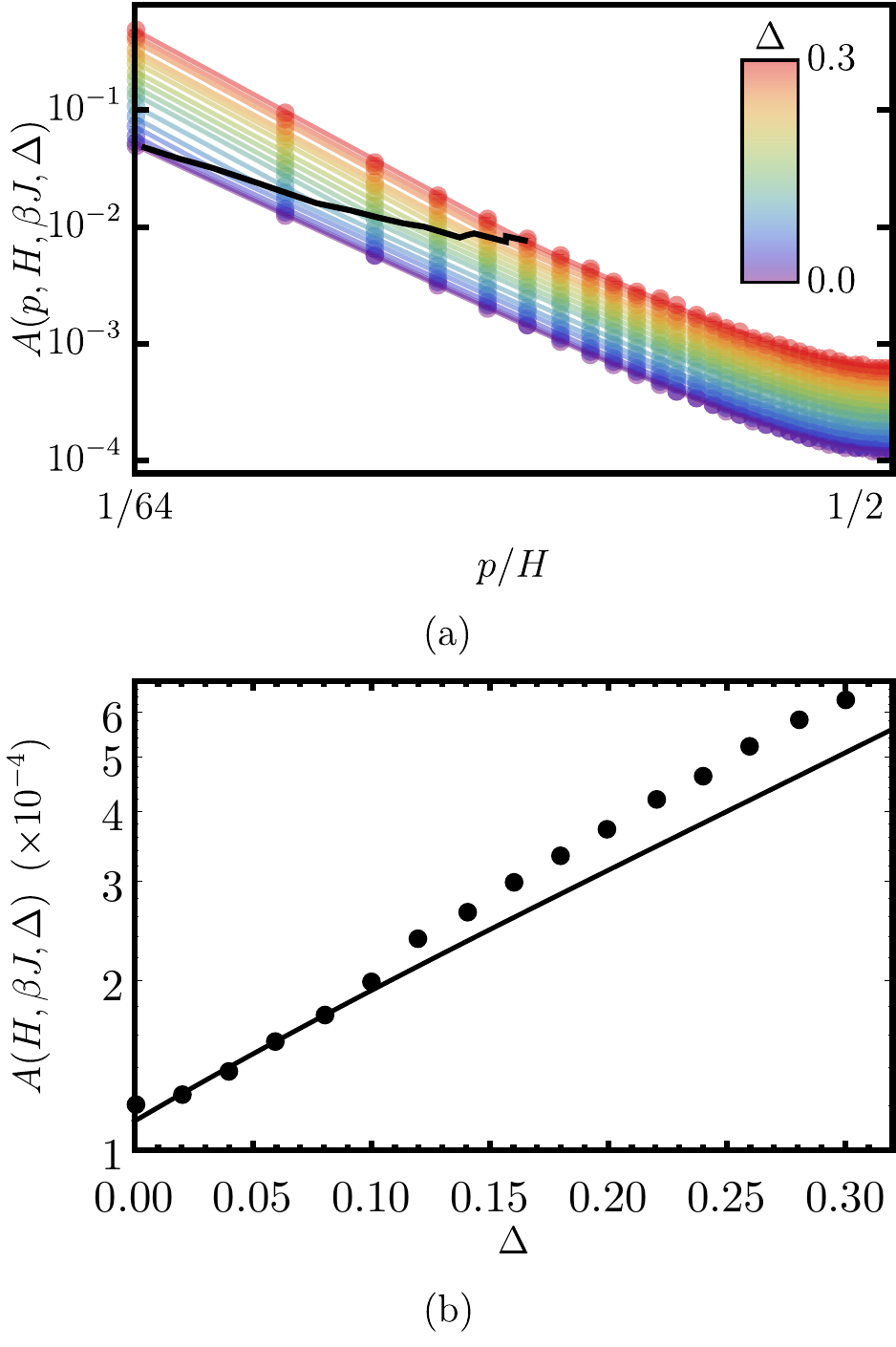}
	\caption{In (a) best values for $A(p,H,\beta J,\Delta)$ together with best fits Eq.~\ref{eq:ap} are shown as functions of $p/H$ for systems of height $H=64$ at temperature $\beta=2.0$ and $0.0\leq \Delta \leq 0.3$.  The black line gives an indication of the crossover point $p_c/H$ at which the system transitions from disorder to thermal induced roughness. Besides the different roughness exponent at large length scales we also observe an overall increase in roughness when more disorder is added to the system. The pre factor $A(H,\beta J,\Delta)$ from Eq.~\ref{eq:ap} is shown as a function of disorder $\Delta$ in (b) together with $A(H,\beta J (1-\Delta))$ as defined in Eq.~\ref{eq:a} and seems to increase exponentially as $\Delta$ increases.
	}
	\label{fig:varyDelta}
\end{figure}

Next we fix the height of the system $H=64$ as well as the temperature $\beta=2.0$ while varying the amount of disorder $\Delta=0.00,0.02,0.04,\dots,0.30$. The increase of disorder leads to a decrease in Larkin length and consequently an increase in where the transition $p_c/H$ occurs. Additionally, the overal pre factor $A(H,\beta J,\Delta)$ increases and thus an intuitive collapse for all $p/H$ is not possible. Instead Fig.~\ref{fig:varyDelta}(a) displays $A(H,\beta J,\Delta)$ as a function of $p/H$ for all values of $\Delta$ at which simulations were performed. The solid black line goes through $p_c/H$ at each value for disorder and thus indicates the crossover point. The pre factor $A(H,\beta J,\Delta)$ is shown as a function of disorder in Fig.~\ref{fig:varyDelta}(b) together with $A(H,\beta J (1-\Delta))$ as defined in Eq.~\ref{eq:a} which suggests an exponential increase in roughness as disorder is increased. 

\begin{figure}\centering
	\includegraphics[width=0.95\linewidth]{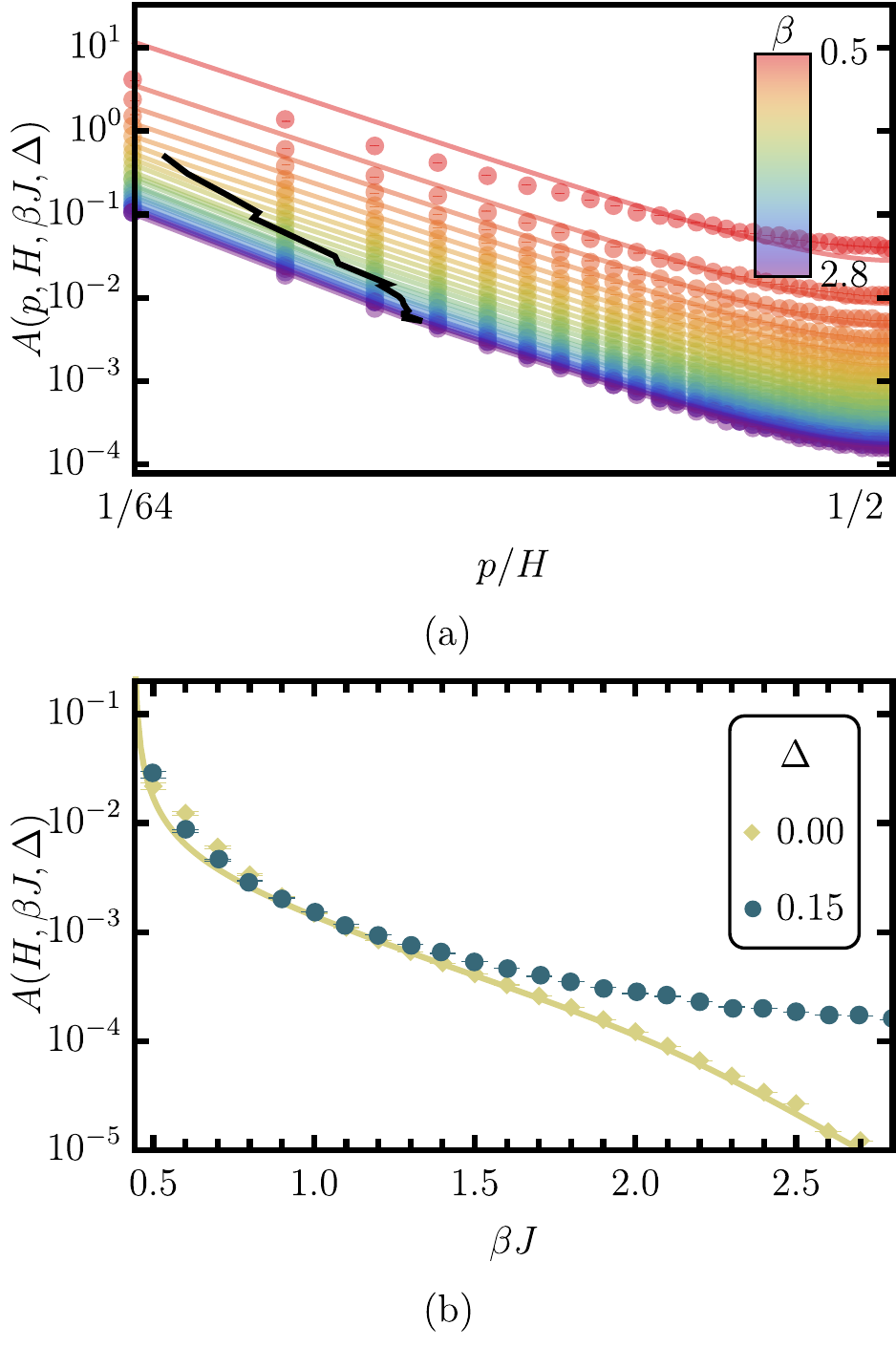}
	\caption{In (a) best values for $A(p,H,\beta J,\Delta)$ and best fits to Eq.~\ref{eq:ap} are shown as functions of $p/H$ for systems of height $H=64$ at $\Delta=0.15$ and temperature $0.5\leq \beta \leq 2.8$. The black line gives an indication of the crossover point $p_c/H$ at which the system transitions from disorder to thermal induced roughness. Note that the crossover point $p_c/H$ seems to have a maximum and will not increase temperature is further decreased beyond a certain point. The pre factor $A(H,\beta J,\Delta)$ is shown as a function of temperature in (b) for $\Delta=0.00$ (yellow) and $\Delta=0.15$ (blue). The theoretical behaviour for a wall in the absence of disorder Eqs.~(\ref{eq:a},\ref{eq:ahigh2}) is shown as the solid yellow line. From both figures we can observe a minimal disorder-induced roughness that endures even in the zero temperature limit.}
	\label{fig:varyBeta}
\end{figure}

We also performed measurements at varying temperatures $\beta=0.5,0.6,\dots,2.8$ at height $H=64$ and $\Delta=0.15$. Figure~\ref{fig:varyBeta}(a) shows $A(p,H,\beta J,\Delta)$ together with a best fit to Eq.~\ref{eq:ap} as a function of $p/H$, together with the transition points denoted by the solid black line. As in the absence of disorder we observe an increase in roughness as temperature is increased. We compare the pre factors $A(H,\beta J,\Delta)$ Fig.~\ref{fig:varyBeta} similar to Fig.~\ref{fig:theoreticalWall}(a) but now for systems at equal height at $\Delta=0.0$ and $\Delta=0.15$. At high temperatures the thermal fluctuations overshadow the effects of disorder and so no difference is observed for $A(H,\beta J,\Delta)$ and Eqs.~(\ref{eq:a},\ref{eq:ahigh2}) still hold. When temperature is lowered, however, disorder plays a more prominent role and $A(H,\beta J,\Delta)$ seems to plateau to the case without disorder. 

We find good agreement between the data from our simulations using the weighed-loop algorithm and Eqs.~(\ref{eq:a},\ref{eq:ahigh2} for large systems without disorder. The introduction of disorder to the system leads to different behaviour at different length scales. At lower length scales, in the thermal regime, $A(p,H,\beta J,\Delta)\sim\csc^2(\pi p/H)\sim p^{-2}$ still holds even for small systems or when there is disorder in the system. As expected we find that the roughness is increased when temperature is increased, larger systems are studied, or disorder is increased. When there is disorder, however, the roughness behaves differently at large length scales $A(p,H,\beta J,\Delta)\sim p^{-7/3}$ which we could also verify with our simulations. The typical length scale $L_c$, at which the roughness depends equally on disorder and thermal fluctuations, decreases when disorder is increased or temperature is decreased but is independent on the size of the system.

\section{Discussion and Conclusion} \label{sec:rbim5}
In this paper we have presented the weighed-loop algorithm that can be implemented for any graph-type model for which the energy only depends on the (one-dimensional) edge of a cluster. Here, we applied it to the square-lattice random bond Ising model (RBIM). More specifically, we have investigated the typical roughness of one-dimensional domain walls in the RBIM. The basic principle of the weighed-loop algorithm is that a weighed directed random walk is performed on a graph. Instead of weighing each possible direction equally we choose to favour directions that are energetically favourable when chosen. This increases the acceptance of proposed updates and thus decreases correlation times especially at lower temperatures or when there is disorder in the system.

The flexibility of the weighed-loop algorithm lies in the function that determines the weights. Throughout this paper we have chosen a function that only depends on the energy contribution of the optional bonds. However, one could also induce a relative or absolute preferred direction by adjusting the weights of optional bonds in different directions. Increasing the odds on straight paths would increase the area of the enclosed cluster for example. It would be interesting to investigate the effects of different weight-functions on correlation times. Additionally, the algorithm can be used in combination with other update schemes and parallel tempering to reduce thermalization. 

Next we compared the weighed-loop algorithm to one of many popular algorithms, Niedermayer's algorithm~\cite{Nie_88}, used to study spin glassy systems. We compared the autocorrelation functions of the square amplitude of the first Fourier mode for domain walls in the RBIM for the two algorithms. The weighed-loop algorithm outperforms Niedermayer's algorithm at low temperature high disorder; the part of parameter space at which many issues arise due to the rough energy landscape and the exponentially high energy barriers. 

To provide a good description for all temperatures below the critical temperature for the roughness of domain walls in the absence of disorder we conciliated the high-temperature expression with our low-temperature expansion. We show good agreement with data obtained with the weighed-loop algorithm and the description in Eqs.~(\ref{eq:a},\ref{eq:ahigh2}).

Once disorder is introduced to the system there appears to be a minimal disorder-induced roughness even when temperature tends to zero. We have investigated the effects of the amount of disorder, temperature, and system size independently. We find that the overall roughness of domain-walls increases as temperature, disorder, or system size is increased.
Besides the change in overall roughness we also observe different behaviour of the roughness at different length scales conform the known theory. For low modes, or equivalently lengths larger than the Larkin length $L_c$~\cite{FGL_89}, the roughness exponent $\zeta_{\text{RB}}=2/3$ is different from the exponent at higher modes $\zeta_{\text{T}}=1/2$. The cross-over point $p_c/H$, at which the transition from disorder-induced to thermal-induced roughness takes place, depend on the temperature and amount of disorder in the system. It increases as disorder is increased or temperature is lowered. In the future we would like to explore this more quantitatively to accurately describe how $L_c$ depends on these parameters, both with simulations and theoretically.


\section{Acknowledgements} \label{sec:rbim6}
We thank E. van der Bijl and G. T. Barkema for valuable feedback, useful discussions, and insights at the early stages of this work. RK is grateful to the Institute for Theoretical Physics at Utrecht University and the Lorentz Institute for Theoretical Physics at Leiden University for the hospitality during the course of this work. This work is part of the D-ITP consortium, a program of the Netherlands Organisation for Scientific Research (NWO) that is funded by the Dutch Ministry of Education, Culture and Science (OCW).

\bibliography{referencesRBIM20170518}

\end{document}